\documentclass[pra,twocolumn,floatfix,a4paper,superscriptaddress,nobalancelastpage]{revtex4}
\usepackage{amsmath,amsfonts,amssymb,graphics,graphicx,epsfig,color,bbm}
\usepackage{times}

\begin{document}
\title{Topological transitions of interacting bosons in one-dimensional bi-chromatic optical lattices}

\author{Xiaolong \surname{Deng}}
\affiliation{Institut f\"ur Theoretische Physik, Leibniz Universit\"at Hannover, Appelstr. 2, 30167 Hannover, Germany}
\author{Luis \surname{Santos}}
\affiliation{Institut f\"ur Theoretische Physik, Leibniz Universit\"at Hannover, Appelstr. 2, 30167 Hannover, Germany}

\begin{abstract}
Ultra-cold atoms in 1D bi-chromatic optical lattices constitute a surprisingly simple system for the study of topological 
insulators. We show that bosons in 1D bi-chromatic lattices present as a general feature the existence at equal fractional filling of 
Mott-insulator phases with different topological character. These different phases are a direct consequence of the 
bosonic and interacting nature of the particles and the topological nature of the Bloch bands.
We demonstrate that the associated hidden topological transitions 
may occur both as a function of the superlattice strength and due to inter-site interactions.
We discuss in addition the topological character of incommensurate density wave phases in 
quasi-periodic superlattices.
\end{abstract}
\date{\today}
\maketitle

\section{Introduction}

Topological insulators~(TIs) are many-body systems that are bulk-gapped, but present topologically protected gapless edge excitations~\cite{Hasan2010, Qi2011}. 
Prominent examples of TIs are provided by the integer quantum Hall effect of 2D electrons in a magnetic field~\cite{Klitzing1980,Prage1987} and 
the quantum spin Hall effect in materials with strong spin-orbit coupling~\cite{Kane2005}. 
TIs have attracted a large deal of attention both due to their fundamental interest and
their applications for quantum computing and spintronics~\cite{Nayak2008, Moore2010}.
TIs of non-interacting fermions may be classified according to their symmetry properties~\cite{Schneider2008,Kitaev2009}.

Symmetry-protected topological~(SPT) phases generalize the idea of TI to 
interacting particles~\cite{Chen2012}. SPT phases are bulk-gapped  and possess robust edge modes as long as the relevant symmetry is not broken. 
Interacting TIs have attracted a large deal of attention~\cite{Raghu2008,Dzero2010,Senthil2013}.
Special interest has been devoted in interacting systems 
to topological phase transitions, i.e. transitions between phases of different topology~\cite{Rachel2010,Wen2010,Varney2010,Jiang2010,Yu2011,Huber2011}.

Atoms in optical lattices~(OLs) constitute an optimal environment for investigating many-body physics~\cite{Lewenstein2007,Bloch2008}. 
In particular, the precise control of the lattice geometry and the inter-particle interactions, and the creation of artificial magnetism for 
neutral atoms~\cite{Dalibard2011}, have opened exciting perspectives for the realization of TIs with cold gases in OLs~\cite{Ruostekoski2002,Hafezi2007,Cooper2008,Palmer2008,Umucalilar2008,Shao2008,Stanescu2010,Goldman2010,Beri2011,Zhang2011,Klinovaja2013}.

Interestingly, a recent experiment 
has shown that 1D photonic quasicrystals may be assigned 
2D Chern numbers, and present topologically protected edge states~\cite{Kraus2012}.
Quasi-periodic 1D potentials may be also realized using incommensurate bi-chromatic OLs~\cite{Roati2008}. Atoms in quasi-periodic OLs constitute hence a surprisingly 
simple scenario for TIs~\cite{Lang2012,Mei2012,Ganeshan2013,Xu2013a,Xu2013b}. Recently it has been shown that a
topological Mott-insulator~(MI) may be realized at particular fractional fillings
in Bose gases in 1D commensurate bi-chromatic OLs~\cite{Zhu2013,Grusdt2013}.

In this paper, we show that bosons in 1D bi-chromatic OLs present as a general feature the existence, at constant fractional filling,  
of various Mott phases that differ in their topological nature. These phases result from the interplay 
between lattice geometry, interactions, and the bosonic nature of the particles. We show that the associated hidden topological transitions 
may occur both as a function of the strength of the secondary lattice, and due to inter-site interactions.
We discuss in addition the topological character of incommensurate density wave phases in 
quasi-periodic bi-chromatic OLs.

The structure of the paper is as follows. In Sec.~\ref{sec:Model} we present the physical system, whereas in Sec.~\ref{sec:Topological} we 
borrow concepts of 2D TI to analyze the effective topological nature of bosons in bichromatic lattices. Sec.~\ref{sec:Quasiperiodic} is devoted to 
quasi-periodic lattices. In Sec.~\ref{sec:Transitions} we reveal hidden topological transitions in 1D bosonic Mott phases as a function of the 
strength of the secondary lattice, whereas in Sec.~\ref{sec:Intersite} we show that such transitions may occur as well due to inter-site interactions. 
Finally, we sum up our conclusions in Sec.~\ref{sec:Conclusions}.


\section{Model}
\label{sec:Model}
We consider a Bose gas loaded in the lowest band of a 1D OL of lattice spacing $\lambda$ with a superimposed lattice of spacing $\lambda/\beta$.
The system is described by the Hamiltonian
\begin{equation}
\!\!H\!=\! -t\!\!\sum_{<i,j>} \!\!\!b^{\dagger}_i b_{j} \!+\! \sum_i \!\left [\frac{U}{2}n_i(n_i \!-\!1)\! +\! \Delta\!\cos\!{(2\pi\beta i \! +\! \phi)} n_i\!\right ]\!\!,
\label{eq:BHH}
\end{equation}
where $<.>$ denotes nearest neighbors, $b^{\dagger}_i$ and $b_i$ are the creation/annihilation operators for bosons at site $i$, $n_i=b^{\dagger}_ib_i$, 
$t$ is the hopping amplitude, $\phi$ is the relative displacement between the lattices, 
$\Delta$ is the strength of the second lattice, and $U$ characterizes the on-site interactions.
We determine the ground-state properties, for both open and periodic boundary conditions~(OBC and PBC), by means of the density matrix 
renormalization group~(DMRG) method~\cite{Schollwock2011}, 
assuming a maximal occupation $n_{max}$ that we increase until reaching convergence.

\section{Effective topological nature} 
\label{sec:Topological}
The effective topological nature of non-interacting fermions in 1D superlattices results from the link to 
electrons in a square lattice with a perpendicular magnetic field~\cite{Lang2012,Hofstadter1976}, which are described by 
the Hamiltonian $H_{2D}=-\sum_{<i,j>} t_{ij}f_i^\dag f_j e^{i2\pi\phi_{ij}}$, where $t_{x,y}$ is the hopping along $x$ and $y$, 
and the magnetic flux through a plaquette is $\Phi=\sum_{\rm plaquette}\phi_{ij}$. Employing Landau gauge, 
the eigenenergies of $H_{2D}$ are given by the 1D model above, with $t=t_x$, $\beta=\Phi$, $\Delta=-2t_y$ and $\phi=-k_y$. 
The effective topological nature may be extended to strongly-interacting 1D bosons, due to the mapping between hard-core bosons~(HCB) and 
non-interacting fermions. We borrow below concepts of 2D TIs to reveal hidden topological transitions in MI phases. 
We note, however, that our results cannot be extrapolated to interacting 2D bosons, where Bose-Fermi mapping is absent.

We impose twisted PBC, introducing a twist angle $\theta\in[0,2\pi)$ 
by considering a tunneling $te^{i\theta/L}$, with $L$ the number of lattice sites.
The topological character of the phase may be discussed using the effective 2D Brillouin zone spanned by the angles $\phi$ and $\theta$.
We may determine the Chern number for the many-body interacting system, following Refs.~\cite{Fukui2005,Varney2011}, 
$C=\frac{1}{2\pi}\int{{\rm d}\theta{\rm d}\phi \mathcal{F}(\theta,\phi)}$, where 
$\mathcal{F}(\theta,\phi)={\rm Im}\left(\left\langle \partial_{\theta}\Psi^{\ast}|\partial_{\phi}\Psi\right\rangle-
\left\langle \partial_{\phi}\Psi^{\ast}|\partial_{\theta}\Psi\right\rangle\right)$ is the Berry curvature~\cite{Niu1985}, 
and $\Psi$ is the many-body ground-state wavefunction. Moreover, we define a phase as \emph{bulk-gapped} if under twisted PBC 
the many-body ground-state remains gapped for all $(\phi,\theta)$ values.

We also consider below OBC for the analysis of 
edge states. For OBC the role of $\phi$ is crucial~(note that $\theta$ is just defined under twisted PBC), because a 
bulk-gapped phase~(in the sense discussed above) may be gapless in the presence of OBC for one or more specific values of $\phi$. In particular, 
borrowing from the bulk-boundary correspondence of 2D TIs, we expect that a bulk-gapped phase with an effective Chern number $C$ 
should present $|C|$ values of $\phi$ at which it becomes gapless under OBC. We show below that this is indeed the case, and that 
the gapless modes correspond indeed to edge states.


\begin{figure} [t]
\resizebox{3.3in}{!}{\includegraphics{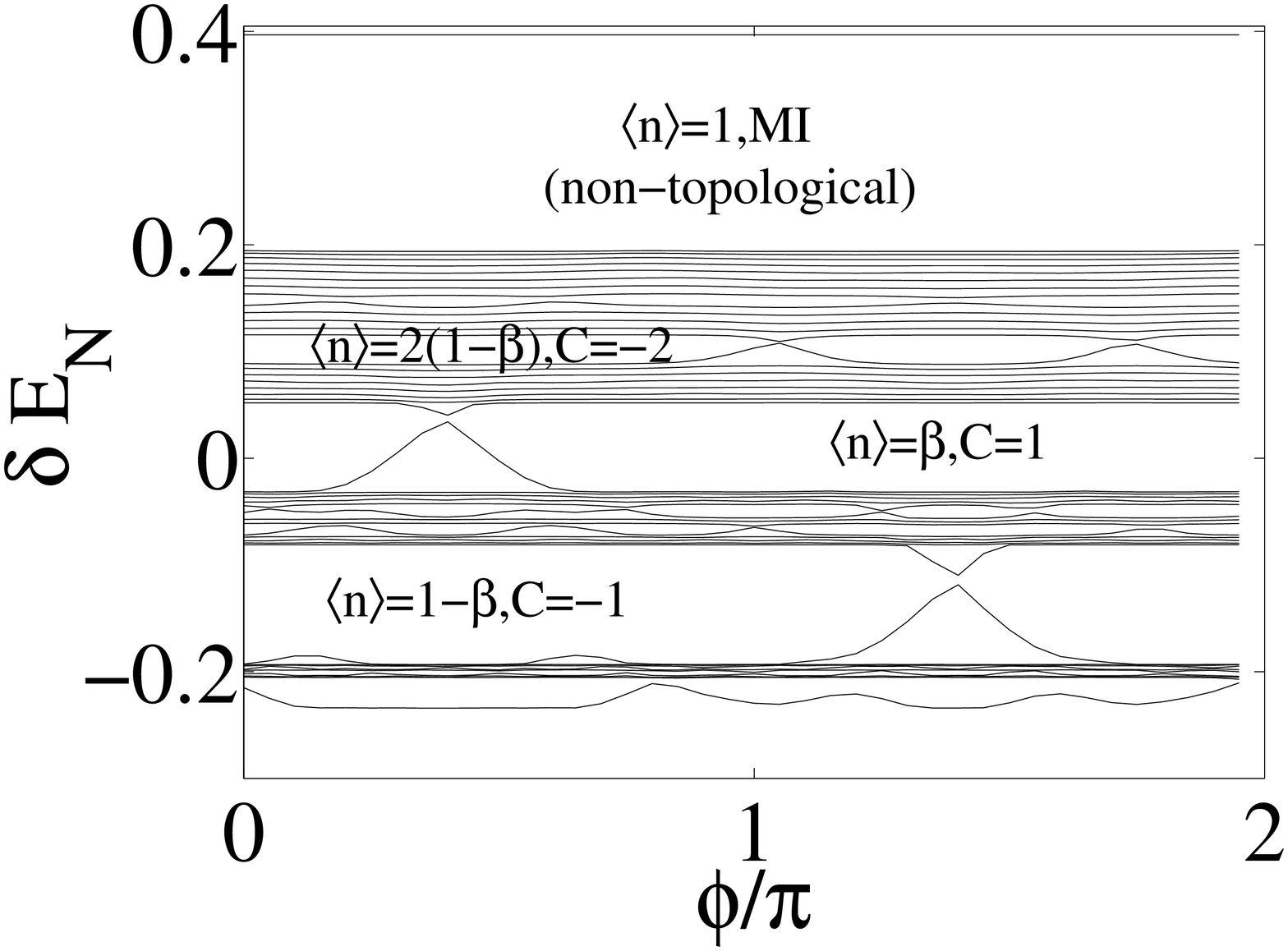}}
\vspace*{-0.3cm}
\caption{Chemical potential $\delta E_N$ of a quasi-periodic model with OBC, for $L=55$ sites, with $U/t=8$ and $\Delta/t=2$.}
\label{fig:1}
\end{figure}


\section{Quasi-periodic potential}
\label{sec:Quasiperiodic}
We start our discussion with the case of incommensurate lattices, which, while being interesting in itself,  
allow us to introduce some key concepts. We consider the particular case of $\beta=(\sqrt{5}-1)/2\simeq 0.618$. 
For large-enough $\Delta$ the system experiences a transition from a superfluid~(SF) phase to a gapless insulating phase 
known as Bose-glass~(BG). Interestingly, for intermediate $\Delta$ values and specific filling factors, as  $\langle n \rangle =\beta$ and $1-\beta$, 
the system becomes a gapped insulator, known as incommensurate charge-density wave~(ICDW)~\cite{Roscilde2008, Roux2008}. 
As shown below, ICDW phases present a non-trivial topological character.

An important insight is obtained from the analysis of the chemical potential at a given particle number $N$, $\delta E_{N}=E_{N+1}-E_N$. 
Figure~\ref{fig:1} shows $\delta E_N$ for $L=55$~\cite{footnote0} and $0\le \langle n\rangle \le 1$ as a function of 
 $\phi$ for $U/t=8$, $\Delta/t=2$, and OBC. Note that the structure of $\delta E_N$ resembles the single-particle spectrum of 
 a TI~\cite{Hasan2010, Qi2011}. In particular, we may define the particle-hole gap 
 $\Delta E_{p-h}(N)=\delta E_{N} - \delta E_{N-1}$, which is particularly clear for ICDW phases 
 with $\langle n \rangle=1-\beta$~($N=21$) and $\beta$~($N=34$).

\begin{figure} [t]
\resizebox{3in}{!}{\includegraphics{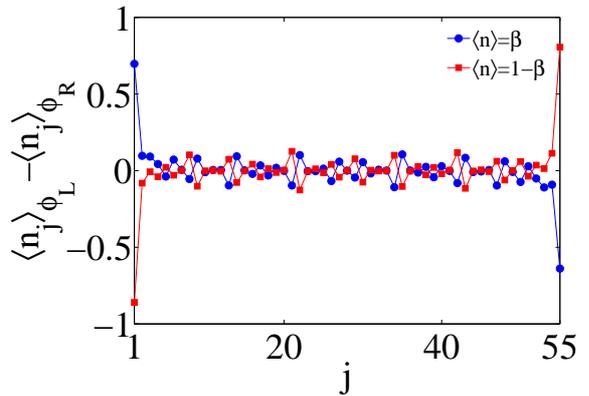}}
\vspace*{-0.3cm}
\caption{Difference in the density distribution for $\phi=\phi_L$~($\phi_R$), slightly smaller~(larger) than the $\phi_{cr}$ value at which $\Delta E_{p-h}$ 
closes for OBC, $L=55$, $U/t=8$, $\Delta/t=2$, and $\langle n \rangle=\beta$ and $1-\beta$.}
\label{fig:2}
\end{figure}

Whereas the ICDW is bulk-gapped, for OBC it becomes gapless for some $\phi=\phi_{cr}$, 
closely resembling the appearance of edge states in the single-particle spectrum of TIs. 
We illustrate in Fig.~\ref{fig:2} the edge-like nature of the in-gap states of Fig.~\ref{fig:1} for the case of $L=55$ with OBC and 
$\langle n \rangle \simeq \beta$ and $1-\beta$. For OBC, close to $\phi_{cr}$, the ground-state and the first excited state are clearly 
separated from other states by the bulk gap. At $\phi=\phi_{cr}$ these two states swap their order. 
Figure~\ref{fig:2} shows the difference between the density profile for $\phi_L$, 
slightly smaller than $\phi_{cr}$, and $\phi_R$, slightly larger than that value. Note that the change in the 
density profile is overwhelmingly concentrated at the edges, showing that 
the ground-state and first excited states just differ by edge excitations. This edge-like nature of 
in-gap states has been observed for all topological phases discussed below.

The non-trivial topology of the ICDW phases is confirmed by calculating the effective Chern number in the way discussed above. 
We obtain in particular $C=-1$ for $\langle n \rangle =1-\beta$ and $C=+1$ for $\langle n \rangle =\beta$~(other ICDW phases are also topological, e.g. for  $\langle n \rangle =2(1-\beta)$ we obtain $C=-2$). This topological character is maintained as long as the system remains in the gapped ICDW phase. 


The $\langle n \rangle=1$ case demands a separate discussion. Two different bulk-gapped phases are possible, 
a MI for $\Delta<U$ and a generalized ICDW~(gICDW) for $\Delta>U$ up to the BG phase~\cite{Deng2012}. The gICDW is characterized by a density 
modulation at the beating quasi-period $1/(1-\beta)$ that is absent in the MI. Under OBC, whereas for the MI $\Delta E_{p-h}$
never closes being $\phi$-independent~(Fig.~\ref{fig:1}), for the gICDW it closes at two different $\phi$ values, which 
are exactly the one for $\langle n \rangle=\beta$ and that for $\langle n \rangle=1-\beta$. 
As for other ICDW phases, the closing of $\Delta E_{p-h}$ for OBC is linked to edge excitations. 
However, the effective Chern number of the gICDW phase is $C=0$. The latter can be intuitively understood as follows.
By comparing the density profile of the gICDW and the profiles for $\langle n\rangle=\beta$ and $1-\beta$, we find that the case 
$\langle n\rangle=1$ can be considered a superposition of two ICDW phases with $\langle n\rangle =\beta$ and $\langle n\rangle=1-\beta$ 
(note also the above mentioned $\phi$ values at which $\Delta E_{p-h}$ closes). 
Recall that these two phases are characterized, respectively, by $C=1$ and $C=-1$, and hence the overall Chern number vanishes. Considering these 
sub-systems as a two-component boson system, we have the equivalent of pseudo-spin-$1/2$ bosons with counter-propagating edge states, reminiscent 
to the case recently discussed in the context of the IQHE for bosons~\cite{Senthil2013}. There is hence a change of the effective topological nature of 
the bulk-gapped phase when growing $\Delta/U$ at the MI-gICDW transition.

\begin{figure} [t]
\resizebox{3.3in}{!}{\includegraphics{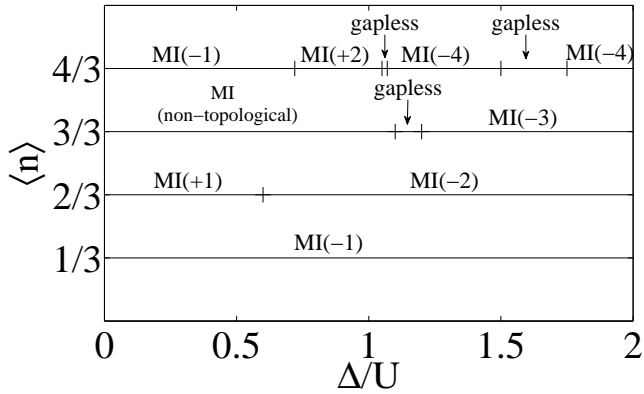}}
\caption{Mott phases for $\beta=1/3$ and $U/t=10$, and $\langle n \rangle=1/3$ to $4/3$. The notation MI(C) denotes a MI phase with Chern number $C$.}
\label{fig:3}
\end{figure}

\section{Topological phase transitions as a function of $\Delta/U$} 
\label{sec:Transitions}
We move at this point to commensurate superlattices, considering the particular example of $\beta=1/3$.
Due to the trimerization of the lattice the lowest energy band splits into three sub-bands.
At a filling $\langle n \rangle=1/3$ and for sufficiently strong interactions a MI occurs with one particle every super-site. 
For HCB~($U\gg t,\Delta$), the 1D bosons may be mapped to free fermions, and the MI may be understood as a band  
insulator with a full lowest sub-band. As a result the MI of hard-core bosons is a TI with Chern number $C=-1$~(we call this phase MI(-1)). 
Due to the gapped nature of the TI, the topological character of the MI is kept as well at finite $U$~\cite{Zhu2013}.
The topology of the MI is maintained for any ratio $\Delta/U$.

A filling $\langle n \rangle =2/3$ results as well in a MI with two particles every super-site. Again, the hard-core case may be mapped to a band insulator 
with each state of the two lowest sub-bands filled once. As a result the MI phase is a TI with Chern number $C=1$~(MI(+1)). As for $\langle n \rangle =1/3$ the 
topological nature persists at finite $U$, but as shown below contrary to the case of $1/3$ filling the behavior of the topological nature of the MI 
for a growing ratio $\Delta/U$ is crucially different.

Figure~\ref{fig:3} shows as a function of $\Delta/U$ our results for $U/t=10$ 
for phases with rational fillings $1/3$, $2/3$, $1$ and $4/3$. 
We have determined for all bulk-gapped phases the effective Chern number using the method described above. 
Topological phases are bulk-gapped but present gapless 
edge states for specific $\phi$ values in the presence of OBC. 
As mentioned above, the bulk-boundary correspondence~\cite{Hasan2010,Qi2011} determines that 
a phase with Chern number $C$ presents $|C|$ edge states, 
i.e. there are $|C|$ different $\phi$ values at which $\Delta E_{p-h}$ closes for OBC.

For $\langle n \rangle =2/3$, as expected from the 
hard-core results, we find a MI(+1) phase for $\Delta/U<(\Delta/U)_{cr}\simeq 0.6$. However, for $\Delta/U>(\Delta/U)_{cr}$ we find 
a different Mott phase~(MI(-2)), again with two particles per super-site, but with $C=-2$. There is hence a topological transition at $(\Delta/U)_{cr}$ linked 
to the closing of the bulk gap at
$(\phi,\theta)=(\pi/3+2n\pi/3,\pi)$ with $n=0,1,2$. The different nature of MI(+1) and MI(-2) is evident from the inset of Fig.~\ref{fig:4}, which shows that with OBC
$\Delta E_{p-h}$ closes for MI(+1) only at a single $\phi$ value, whereas MI(-2) closes at two values (and hence has two associated edge-like excitations), 
in accordance to the bulk-boundary correspondence.

We stress that although the topological character of MI(+1) may be well understood from the hard-core limit, the topological transition into MI(-2) 
is a direct consequence of the bosonic and interacting nature of the particles. 
The physics behind the different topological phases may be easily illustrated using the case $\phi=\pi$. When $\Delta$ grows the separation between 
sub-bands increases and each sub-band becomes narrower. The phase transition is linked to the fact that for $\beta=1/3$ and 
$\langle n \rangle = 2/3$ an insulator may occur either by filling with one particle each Bloch state of the lowest two sub-bands, or by filling twice 
each Bloch state of the lowest sub-band. Whereas this observation is inconsequential for free fermions or HCB, it becomes crucial for interacting soft-core bosons. 
For large $\Delta\gg U$, the system minimizes the energy by placing all particles into the lowest sub-band, which is possible 
only due to the bosonic nature of the system. Moreover, due to the very narrow 
bandwidth (much smaller than $U$) of the lowest sub-band for $\Delta\gg U\gg t$, the Bose gas minimizes its interaction energy by populating each available Bloch state of the lowest sub-band twice. 
As a consequence, the system behaves as two exact copies of the lowest sub-band, and hence the system is an insulator with 
a Chern number twice that of the lowest sub-band, i.e. $C=-2$.

A similar reasoning may be applied for higher fillings, explaining the results of Fig.~\ref{fig:3}. In particular, whereas in the hard-core regime 
a filling $\langle n \rangle=1$ results  
in a non-topological MI~(with all Bloch states of the three sub-bands filled once), for large $\Delta/U$, the system minimizes 
the energy by filling three times each state of the lowest sub-band. 
As a result we expect a topological phase with $C=-3$ at large $\Delta/U$, as we have confirmed in our DMRG calculations.
Finally, for $\langle n \rangle = 4/3$, we expect $C=-1$ for small $\Delta/U$ since the energy is minimized by filling 
the upper sub-bands once and the lowest one twice~(and hence $C=-1$).  
At large $\Delta/U$ the lowest sub-band will be four-fold filled, and hence we expect $C=-4$ as observed in our numerical results. 
Moreover, a third bulk-gapped topological phase occurs with $C=+2$, 
corresponding to a double filling of the two lowest sub-bands.

The same arguments may be applied to Mott phases for bosons in superlattices with other commensurate values of $\beta$. 
Hence, the appearance of Mott phases with different topologies as a function of the secondary lattice depth is predicted to be a general feature of 
bosons in 1D superlattices.

\section{Topological phase transitions due to inter-site interactions}
\label{sec:Intersite}
Although in current OL experiments only on-site interactions are relevant, inter-site interactions 
are expected to play a crucial role in future experiments with cold polar molecules, due to the long-range character of the electric 
dipole-dipole interaction~\cite{Baranov2008,Lahaye2009,Baranov2012}. We consider for simplicity only nearest-neighbor~(NN) interactions, 
which result in an additional term $V\sum_i n_i n_{i+1}$ in Eq.~\eqref{eq:BHH}. Inter-site interactions may drive as well topological phase transitions in superlattice 
bosons, as we illustrate below for $\beta=1/3$ and $\langle n \rangle = 2/3$. 


\begin{figure}[t]
\resizebox{3.3in}{!}{\includegraphics{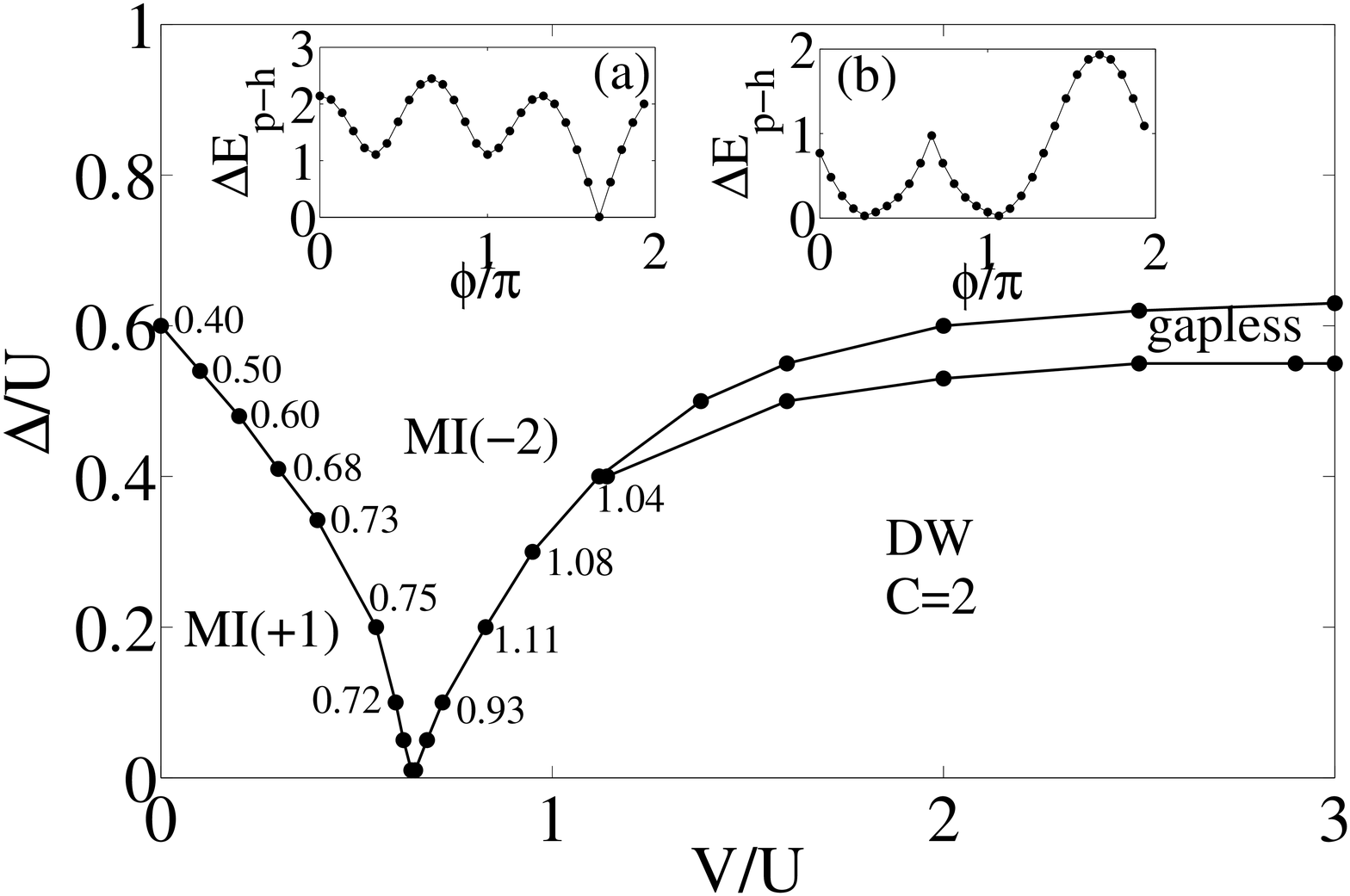}}
\caption{Phase diagram for NN interacting bosons for $\beta=1/3$, $\langle n \rangle=2/3$, and $U/t=10$. 
We indicate the exponent $\alpha$ at the MI(+1)-MI(-2) and MI(-2)-DW transitions~(see text).
The insets show $\Delta E_{p-h}$ as a function of $\phi$ for OBC for $\Delta/U=0.2$ with (a) $V/t=0.1$~(MI(+1)) and (b) $V/t=0.7$~(MI(-2)).
}
\label{fig:4}
\end{figure}


Figure~\ref{fig:4} shows our results for 
the ground-state phases for $U/t=10$. At small $V/U$ and $\Delta/U$~\cite{footnote1} we recover, as expected from the hard-core limit, the MI(+1) phase.
For large $V/U$, the strong NN repulsion results in a doubly-degenerate ground state, characterized by 
a density modulation with a six-site periodicity~\cite{footnote2}. 
Due to this modulation we denote this phase as density-wave~(DW) phase. 
The DW phase is characterized by a Chern number $C=2$ resulting from the sum of the Chern numbers ($C=1$) 
of both degenerate states. Between the MI(+1) and DW phases we find the MI(-2) phase. 
The MI(+1)-MI(-2) transition is marked, as mentioned above, by a vanishing bulk gap for 
$(\phi,\theta)=(\pi/3+n2\pi/3,\pi)$ with $n=0,1,2$. These bulk-gapless points 
are characterized by the non-universal exponent $\alpha$ of the single-particle correlation function 
$\langle b^{\dagger}_0 b_r\rangle \propto r^{-\alpha}$, which we have determined~(Fig.~\ref{fig:4}) 
using conformal scaling~\cite{Roux2008}. 
For low $\Delta/U$ we observe a critical bulk-gapless line separating MI(-2)-DW~(the bulk gap closes for $(\phi,\theta)=(n 2\pi/3,\pi)$, with $n=0,1,2$) 
characterized by $\alpha\simeq 1$. For large $\Delta/U$ 
our numerical results reveal a finite bulk-gapless region separating the MI(-2) and DW phases.\\


\section{Conclusions}
\label{sec:Conclusions}
We have shown that superlattice bosons  present Mott-like phases with the same fractional filling but different topological character. The 
hidden topological transitions at constant filling may occur both as a function of the superlattice strength and due to inter-site interactions.

We note that typical OL experiments present an overall harmonic confinement~(crucial for the existence of Mott phases 
at the required fillings in some spatial regions). For free fermions or hard-core bosons the topological character of the MI phases may be 
evaluated from the inhomogeneous density profile, averaged 
over the elementary cell (of $s$ sites), $\bar n_j=\frac{1}{s}\sum_{i=j}^{j+s-1} \langle n_j \rangle$ using Streda's formula~\cite{Umucalilar2008,Lang2012,Mei2012, Zhu2013,Streda1982}. 
Note, however, that MI(+1) and MI(-2) have the same $\bar n_j$, and hence Streda's formula cannot be employed for revealing the topological transition. 
This change may be however investigated in the presence of sharp boundaries that may be induced in OLs using laser-assisted 
tunneling as discussed in Ref.~\cite{Goldman2010}.  As mentioned above, whereas MI(+1) presents gapless edges around a single 
$\phi$ value, MI(-2) shows edges around two different $\phi$ values. Moreover, the topological transition as a function of $\Delta$ opens the possibility 
of creating a MI(+1)-MI(-2) boundary by employing a spatially varying strength $\Delta$. Since $C$ jumps from $1$ to $-2$, the presence of such a boundary is accompanied by 
edge states at three different $\phi$ values.

\acknowledgements
We thank T. Vekua for helpful discussions. The work is supported by the excellence cluster QUEST.


\begin{thebibliography}{99}
\bibitem{footnote0} $L$ values within the Fibonacci series, as e.g. $L=55$, must be employed for $\beta=(\sqrt{5}-1)/2$ 
when studying PBC. We employ for simplicity the same $L$ for OBC. We stress however, that edge-like states are obtained for OBC for 
general $L$ values, although the $\phi$ value at which $\Delta E_{p-h}$ closes changes.
\bibitem{footnote1} Note that the superlattice strength remains finite, $\Delta/U>0.01$ in Fig.~\ref{fig:4}, since for $\Delta=0$ neither a MI nor a DW are possible, and the system 
is found in a SF phase.
\bibitem{footnote2} E.g., for  $\phi=0$, each three-site elementary cell $j$ has two equal low sites~($1,2$) and an upper one~($3$), and may be characterized by the 
occupation $(n_{j,1},n_{j,2},n_{j,3})$. For $U,\Delta,V\gg t$, the ground state presents a distribution $(0,2,0)$ for even~(odd) $j$ and $(1,0,1)$ for odd~(even) $j$. 
Similar ground states are found for other $\phi$ values.  Note that due to the six-site periodicity of the DW phase, we need to perform calculations for $L$ equal to a multiple of $6$. 
We perform in particular exact diagonalization calculations for $L=6$ and $12$ and DMRG for $L=18$ and $24$.
\end{thebibliography}
\end{document}